\documentclass[12pt]{iopart}
\usepackage{graphicx}
\usepackage{epstopdf}
\usepackage{color}
\usepackage{amsfonts}
\usepackage{mathptmx}
\usepackage[caption = false]{subfig}

\newcommand{\diff}[1]{\mathrm{d}{#1}}
\newcommand{\pder}[2]{\frac{\partial{#1}}{\partial{#2}}}

\newcommand{\Lagr}{\mathcal{L}}
\newcommand{\Unor}{\mathcal{U}}

\newcommand{\sgn}{\textrm{sgn}}

\begin{document}

\title{Extremally charged line}

\author{Ji\v{r}\'{\i} Ryzner \& Martin \v{Z}ofka}

\address{Institute of Theoretical Physics, Charles University in Prague}
\ead{j8.ryzner@gmail.com, zofka@mbox.troja.mff.cuni.cz}
\vspace{10pt}

\begin{abstract}
We investigate the properties of a static, cylindrically symmetric Majumdar-Papapetrou-type solution of Einstein-Maxwell equations. We locate its singularities, establish its algebraic type, find its asymptotic properties and weak-field limit, study the structure of electrogeodesics, and determine the mass and charge of its sources. We provide an interpretation of the spacetime and discuss the parameter appearing in the metric.
\end{abstract}

\pacs{04.40.Nr, 04.20.Jb}

\vspace{2pc}
\noindent{\it Keywords}: Majumdar-Papapetrou spacetime, line singularities, electrogeodesics

\submitto{\CQG}

\maketitle

\section{Introduction}
The Majumdar-Papapetrou (MP) solution \cite{majumdar, papapetrou} representing an arbitrary set of stationary, extremally charged black holes in equilibrium is well known\footnote{The term extremal refers to the fact that the charges of the black holes are equal to their masses, rendering their horizons degenerate.}. The spacetime arises as a solution of Laplace's equation. Hartle and Hawking \cite{hartle} assumed a flat spatial infinity and showed that any solution to Laplace's equation with non point-like sources must contain a naked singularity. MP spacetimes with localized sources are unique in the sense that the only $\mathcal{I}^+$-regular stationary and axisymmetric solutions of Einstein-Maxwell equations are the Kerr-Newman metrics and the MP metrics \cite{Chrusciel_Costa_Heusler} and they can be seen as a solution to the Cauchy initial value problem \cite{Alcubierre_Degollado_Salgado}. Furthermore, the asymptotically flat subset of the Majumdar-Papapetrou class and the Reissner-Nordstr\"{o}m exterior solution are the only asymptotically flat conformastatic electrovacuum spacetimes \cite{Gonzalez_Vera}. The two-hole MP spacetime can also be obtained as the extreme limit of the double RN solution \cite{Cabrera-Munguia_Manko_Ruiz}. Recently, MP spacetimes were used to study the influence of external sources on black holes since one expects that the most powerful effect is to be expected from an additional black hole \cite{Semerak_Basovnik}.

There are, however, interesting classes of solutions of different asymptotics. In this paper, we assume a line source that extends to infinity along a straight line and we thus do not have a flat spatial infinity. Although it does contain a naked singularity it is of interest in itself and, generally, non-asymptotically flat MP solutions may involve horizons. The importance of the solution consists in the fact that it has an obvious classical analog we can compare it to. We investigate the properties of the spacetime and also compare it to the charged black string \cite{lemos}, which however requires a non-zero cosmological constant. We interpret the parameter of the metric using electrogeodesics, integral definitions of mass and charge, and Israel formalism for various shell sources replacing the singularities of the solution. We first derive the physical properties of the non-relativistic counterpart of the solution to be later able to compare it to the full solution. In Chapter 3, we present the studied spacetime and its basic geometrical characteristics. In Chapters 4 and 5, we determine the mass and charge of the singular line sources using integral definitions of energy and the Israel formalism, respectively. In Chapter 6, we finally investigate motion of both charged and uncharged massive particles and photons and compare them to the Newtonian case.
\section{Newtonian analog}
We review the classical analog of the infinite charged string. We first write the gravitational and electrostatic potentials, $\varphi_G$ and $\varphi_E$, in the standard cylindrical coordinates $\rho, \phi, z$ as follows
\begin{equation}
\varphi_G = 2\mu \ln \frac{\rho}{P}, \, \varphi_E = -2\lambda \ln \frac{\rho}{P},
\end{equation}
with $\mu$ the mass and $\lambda$ charge per unit length of the string. $P$ is a nor\-ma\-li\-za\-tion constant defining the cylindrical surface of vanishing potential\footnote{We chose both potentials to vanish at the same radius so that they are proportional to each other; we may always do so as the two constants only appear as a constant in the Lagrangian (\ref{lagrangian}).}. We now rescale the radial and azimuthal coordinates so that $\rho/P \rightarrow \rho, z/P \rightarrow z$. The classical Lagrangian per unit mass of a massive and charged test particle of specific charge $q=Q/M$ then reads
\begin{equation}\label{lagrangian}
\Lagr = \frac{1}{2}\left(\dot{\rho}^2 + \rho ^2 \dot{\phi}^2 + \dot{z}^2\right)-\left(q\varphi_E+\varphi_G\right).
\end{equation}
The Lagrangian does not depend on $\phi$ and $z$ and does not explicitly depend on $t$, so we have the following integrals of motion
\begin{eqnarray}
\label{eq:klasrce1}
E & \equiv & \sum_{i} \pder{\Lagr}{\dot{q}^i}q^i - \Lagr = \frac{1}{2}
\left( \dot{\rho} ^2 + \rho ^2 \dot{\phi}^2+ \dot{z}^2 \right)+ \varphi_G + q \varphi_E,\\
L_z &\equiv & \pder{\Lagr}{\dot{\phi}}=\dot{\phi} \rho ^2,\\
p_z & \equiv & \pder{\Lagr}{\dot{z}} =  \dot{z}.
\end{eqnarray}
The only remaining equation of motion is
\begin{equation}
\label{eq:klasrce4}
2\left(q \lambda - \mu\right) + \rho^2 \dot{\phi}^2 - \rho \ddot{\rho} = 0.
\end{equation}
\subsection{Static solution}
If the test particle is to be in equilibrium at a given point we must have
\begin{equation}
\label{eq:klasstatic}
\mu - q \lambda \equiv \mathcal{A} = 0, E = p_z =  L_z = 0.
\end{equation}
Thus, the particle can only remain at rest if it is made of the same material as the source as regards its specific charge and in such a case it can stay still at any point. In fact, the potential part of the Lagrangian then cancels out and we have a particle moving along straight lines at a constant velocity. Also, free-particle motion is the only case admitting purely axial motion (with $E, p_z \not=0$).
\subsection{Radial motion}
For radial motion, with $z$ and $\phi$ constant, we obtain a single equation
\begin{equation}
E = \frac{1}{2}\left[\dot{\rho}^2+4\mathcal{A}\ln \rho\right],
\end{equation}
which can be rewritten as
\begin{equation}
\dot{\rho}^2 = 2\left[E -2 \mathcal{A} \ln \rho \right].
\end{equation}
Motion is only possible if the right-hand side is non-negative. Omitting the above case of a freely moving particle with $\mathcal{A}=0$, we then have two cases depending on the sign of $\mathcal{A}$: unbound orbits reaching the radial infinity with $\mathcal{A}<0$ and bound orbits intersecting the source otherwise. Motion can be expressed explicitly in terms of the error function.
\subsection{Circular motion}
For circular orbits with $\rho$ and $z$ constant, we find the following equations
\begin{eqnarray}
0 &=&\rho^2 \dot{\phi}^2 -2\mathcal{A}\\
L_z &=& \rho ^2 \dot{\phi}, p_z=0 \label{classical equation for phidot}\\
E &=& \frac{1}{2}\left[\rho^2 \dot{\phi}^2 +4\mathcal{A}\ln \rho\right]
\end{eqnarray}
This yields a constant angular velocity
\begin{equation}
\omega = \frac{\sqrt{2 \mathcal{A}}}{\rho}.
\end{equation}
Circular motion can occur at any radius but only for $\mathcal{A} > 0$, which means that gravity is stronger than the electromagnetic force.
\section{Geometry of the spacetime}
The investigated general relativistic solution stems from the same Laplace equation as the classical field of Chapter 2. Based on this analogy, we shall refer to it as the extremally charged string or ECS, for short. However, the gravitational and electromagnetic fields are not independent here and are given by the same function. One integration constant, $P>0$, can be scaled away by introducing dimensionless cylindrical coordinates, $\rho/P \rightarrow \rho, z/P \rightarrow z, t/P \rightarrow t$. The investigated spacetime is then described by the rescaled metric
\begin{equation}
\mathrm{d}\tilde{s}^2 = \frac{\mathrm{d}s^2}{P^2} = -\frac{\diff{t}^2}{U^2}+U^2 \left(\diff{\rho}^2 + \rho ^2 \diff{\phi}^2 + \diff{z}^2\right),
\end{equation}
with
\begin{equation}
U(\rho) = 1+K \ln \rho,
\end{equation}
where $K$ is the other integration constant the meaning of which is one of our goals. The form of the potential $U$ is chosen in such a way that the limit $K \rightarrow 0$ corresponds to the Minkowski space as discussed below. The electromagnetic four-potential is
\begin{equation}
A = \frac{\diff{t}}{U},
\end{equation}
yielding the following Maxwell tensor
\begin{equation}\label{Maxwell tensor}
  F = \frac{U_{,\rho}}{U^2} \diff{t} \wedge \diff{\rho},
\end{equation}
describing a purely radial electric field $E_\rho = -K/\rho (1+K \ln \rho)^2$, which vanishes at radial infinity. The spacetime is static and cylindrically symmetric, admitting only these three Killing vectors.

The Kretschmann scalar reads
\begin{equation}
R_{\mu \nu \kappa \lambda}R^{\mu \nu \kappa \lambda} = \frac{8 K^2 \left[2 K^2 \ln ^2 \rho +7 K^2+2 (3 K+2) K \ln \rho+6 K+2\right]}{\rho ^4 \left(K \ln \rho +1\right)^8},
\end{equation}
and vanishes far away from the axis, $\rho \rightarrow \infty$, as do also all tetrad components of the Riemann tensor. The spacetime thus has two singularities: one located at $\rho = 0$ while the outer one has $\rho \equiv \rho_o = \exp (-1/K)$. The spacetime thus splits in two independent regions separated by the outer singularity.

The spacetime is generally of type I apart from two special cylindrical surfaces where it is type D. Additionally, at radial infinity, it approaches type O as all Weyl scalars vanish in the limit $\rho \rightarrow \infty$.

Lemos and Zanchin \cite{lemos} found a spacetime also describing the field of a charged and massive infinite straight string. Their solution, however, requires the presence of a negative cosmological constant balancing the field. Therefore, the asymptotics of the solution far away from the axis approach that of anti-de Sitter. The string itself is singular but it always has a horizon making sure the cosmic censorship conjecture holds. The mass and charge densities are independent. These are all points of difference between the two solutions. If there are any closer similarities they might be revealed due to the fact that our solution can also be generalized to contain a positive cosmological constant, which will change its asymptotics and cancel the static nature of the spacetime. In our future work we will study the cosmological solution in more detail.
\subsection{Proper lengths}
\begin{figure}[h]
  \centering
    \includegraphics[width=90mm,angle=0]{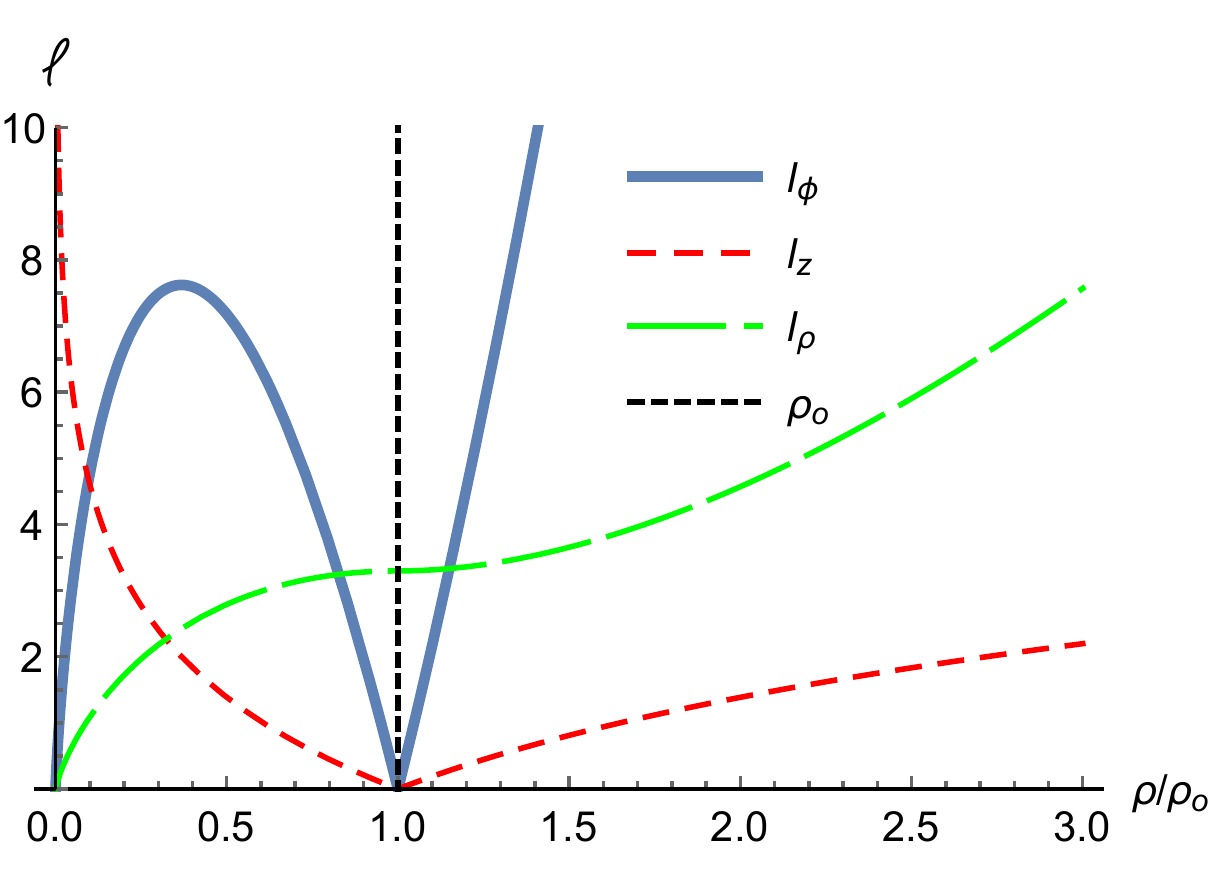}
  \caption{Proper length of various curves for $K = -2$. The form of the curves is the same also for $K>0$.}
  \label{fig:properlength}
\end{figure}

Let us now investigate the proper length of some special curves. Let us begin with the proper circumference of a circle of constant $\rho$
\begin{equation}\label{hoop_circumference}
\diff{l_{\phi}}^2 = \rho^2 U^2 \left(\rho\right) \diff{\phi}^2 \Rightarrow l_{\phi}(\rho) = 2\pi \rho |1+K \ln \rho|.
\end{equation}
The circumference of the hoops vanishes at $\rho=0$ then grows to its maximum value at $\rho_c = \exp(-1 - 1/K) = \rho_0/e < \rho_0$ and vanishes again at $\rho = \rho_0$. The outer singularity thus behaves as another axis of the spacetime, see \Fref{fig:properlength}. The proper length of the coordinate segment $(0,h)$ along the $z$-axis is
\begin{equation}
\diff{l_{z}}^2 = U^2 \left(\rho\right) \diff{z}^2 \Rightarrow l_z(\rho) = h|1+K \ln \rho|.
\end{equation}
The outer singularity thus contracts along its length as well and appears to be a point rather than a cylindrical surface. The proper distance from $\rho=0$ is given by
\begin{equation}
\diff{l_{\rho}}^2 = U^2 \left(\rho\right) \diff{\rho}^2 \Rightarrow l_{\rho}(\rho) = \int_{0}^{\rho} |1+K \ln \rho'| \diff{\rho'}.
\end{equation}
To compute the integral, we need to split the integration into cases when $0~<~\rho~<~\rho_o$ and $\rho \geq \rho_o$. Assuming the non-trivial case of $K \not = 0$, we obtain
\begin{equation}
   l_{\rho}(\rho) = \left\{
     \begin{array}{lr}
       - \rho \left(1-K+K \ln \rho \right) \sgn K, & 0 < \rho < \rho_o,\\
       \left[ \rho \left(1-K+K \ln \rho \right) +2K \rho_o\right] \sgn K , & \rho \geq \rho_o.
     \end{array}
   \right.
\end{equation}

For $K>0$ and in the limit $K \rightarrow 0^+$, the spacetime yields the Minkowski spacetime above $\rho_o$ while the inner region shrinks with the proper distance from the axis to $\rho_o$ vanishing and from $\rho_o$ to infinity diverging as $\rho$. For $K<0$ and in the limit $K \rightarrow 0^-$, we find Minkowski below $\rho_o$ with the proper distance from the axis to $\rho_o$ infinite.
\section{Mass and charge of ECS}
In this section we focus on the total energy enclosed within a static cylinder of constant radius, $\rho$. The advantage of ECS is that it is static and expressed in coordinates where the metric is diagonal. However, ECS is not asymptotically flat so we cannot use, for example, the ADM mass. We thus apply several other definitions of mass (energy) enclosed within a coordinate cylinder and compare them. Finally, we also calculate the charge enclosed in the respective cylinders. Since ECS is a special case of Majumdar-Papapetrou spacetimes with equal linear charge and mass densities we expect a similar behavior at least in the weak-field limit of $K \rightarrow 0$.
\subsection{C-energy}
We cast the metric as
\begin{equation}
\diff{s}^2 = U^{-2}\left(-\diff{t}^2+\diff{R}^2\right)+U^{2}\diff{z}^2+\rho^2 U^{2} \diff{\phi}^2,
\end{equation}
corresponding to the canonical form required to determine the C-energy \cite{thorne}. After regularization, we obtain
\begin{equation}
\mathcal{E}_C \left(\rho\right) = \frac{h}{8} \left[1 -  \left(\frac{U+2\rho U_{,\rho}}{U}\right)^2\right] = -\frac{h}{2} K \frac{1+ K + K \ln \rho}{(1+ K \ln \rho)^2}.
\end{equation}
In the limit $K \rightarrow 0$, we get
\begin{equation}
\mathcal{E}_C \left(\rho\right) \approx -\frac{hK}{2}+\frac{h}{2}\left(-1+\ln \rho\right)K^2 + O\left(K^3\right).
\end{equation}
For a plot of the resulting function, refer to \Fref{fig:summarymasscharts}. It diverges at the outer singularity and vanishes both at $\rho=0$ and at the radial infinity.
\subsection{Landau-Lifshitz}
Landau and Lifshitz derived a conservation law \cite{landaulifschitz}
\begin{equation}\label{eq:landaumass}
\left[16\pi \left(-g\right)\left(T^{\mu \nu}+ t^{\mu \nu}_{LL}\right)\right]_{,\nu} = 0,
\end{equation}
with $g$ the determinant of the metric and based on the stress-–energy pseudotensor of the gravitational field, $t^{\mu \nu}_{LL}$, defined as follows
\begin{equation}
16\pi t^{\mu \nu}_{LL} \equiv g^{-1} \left[g \left(g^{\mu \nu}g^{\alpha \beta} - g^{\mu \alpha}g^{\nu \beta}\right)\right]_{,\alpha \beta} - \left[2R^{\mu \nu}+\left(2 \Lambda -R\right)g^{\mu \nu}\right].
\end{equation}
In our calculation we determined the corresponding super-potential (see \cite{landaulifschitz}) and obtained a relation for the mass
\begin{equation}
M_{LL}(\rho) = P^{0} = \frac{h}{2} |K| \sgn\left(\rho_o-\rho\right)\left(1+K \ln \rho\right)^6.
\end{equation}
For $K \rightarrow 0$, we find
\begin{equation}
M_{LL}(\rho) \approx -\frac{h K}{2} -3 K^2 h \ln \rho +O\left(K^3\right).
\end{equation}
For a plot of the resulting function, refer to \Fref{fig:summarymasscharts} where we can see that $M_{LL}(\rho)$ diverges at $\rho=0$ and radial infinity and vanishes at $\rho=\rho_o$. It changes sign at the outer singularity and the form of its plot is independent of the sign of $K$.
\subsection{Brown-York}
Energy is defined as an integral over the boundary $S$ of a volume $\Sigma$ and one needs to subtract the contribution of a selected background spacetime \cite{brown}. Since in the limit $K \rightarrow 0$ we obtain Minkowski, we subtract the contribution of flat spacetime \cite{lemos} from the general relation
\begin{equation}
M_{BY}  =  \int_{S} g_{\mu \nu}\left(\mathcal{E} n^{\mu} + j^{\mu}\right) \xi_{\left(t\right)}^{\nu} \diff{S}, \label{eq:LemosEnergy}
\end{equation}
with appropriate definition of the terms appearing in the integral (see \cite{lemos}). After some algebra, we conclude
\begin{equation}\label{eq:LemosJakoKomar}
M_{BY}\left(\rho\right) = \frac{-h K}{2\left(1+K \ln \rho\right)}.
\end{equation}
For $K \rightarrow 0$, we find
\begin{equation}
M_{BY}\left(\rho\right) \approx -\frac{hK}{2}+\frac{hK^2}{2} \ln \rho + O\left(K^3\right).
\end{equation}
For a plot of the resulting function, refer to \Fref{fig:summarymasscharts} (denoted $M_K$, see below): the Brown-York mass is positive below the outer singularity and negative above it. It vanishes at $\rho=0$ and at infinity while diverging at $\rho=\rho_o$.
\subsection{Komar mass}
For a stationary spacetime, Komar defines the mass enclosed in a three-dimensional spacelike surface $\Sigma$ \cite{reltoolkit} as
\begin{equation}\label{eq:definiceMK}
M_K = \frac{1}{4\pi} \oint_{S} \xi_{(t)}^{\alpha;\beta} r_{\alpha} n_{\beta} \diff{S}
\end{equation}
where $\xi^{\beta}_{\left(t\right)}$ is the Killing vector corresponding to the time symmetry. Now we plug in our choice of integration surface to yield
\begin{equation}
M_K\left(\rho\right) =   -\frac{h \rho U_{,\rho}}{2 U}.
\end{equation}
We see that this expression is identical to the Brown-York definition (\ref{eq:LemosJakoKomar}).
\begin{figure}[h]
\centering
\includegraphics[width = 90mm]{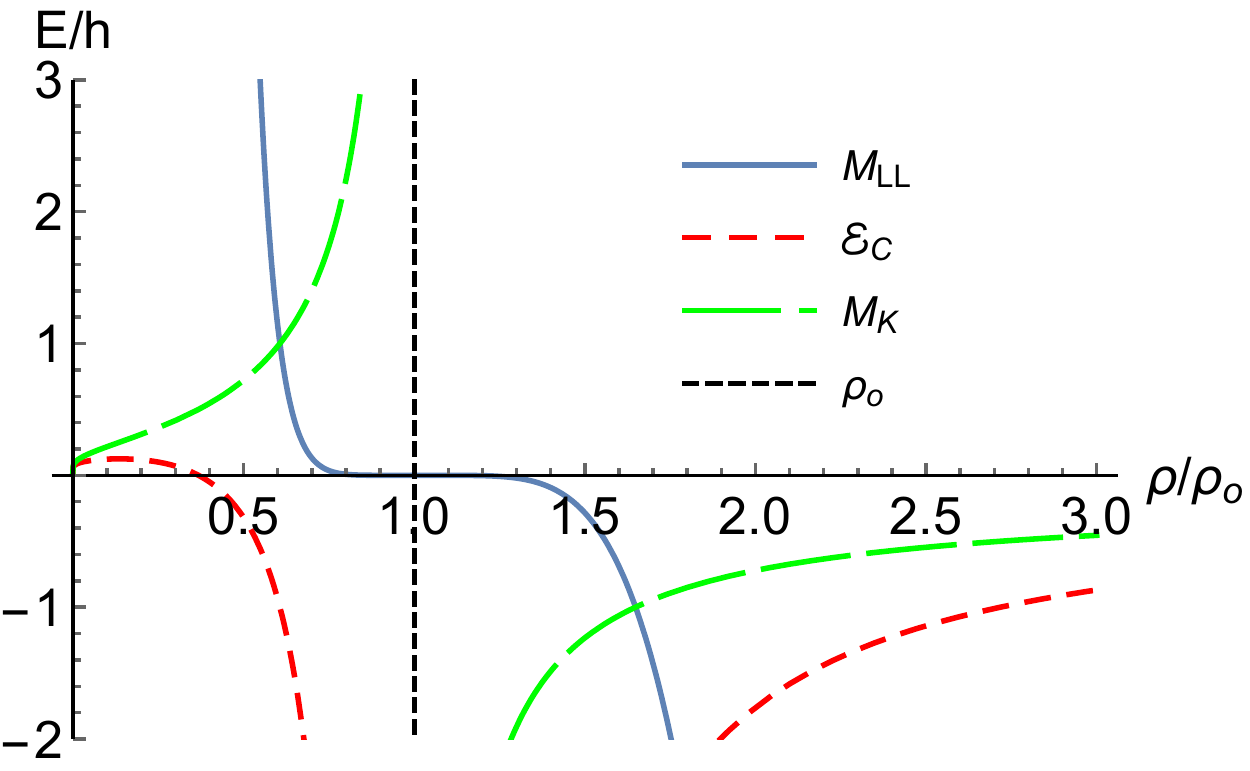}
\caption{Comparison of different definitions of mass for $K=\pm 2$.}
\label{fig:summarymasscharts}
\end{figure}
\subsection{Overview of results for mass}
We summarize our results for various definitions of mass in \Fref{fig:summarymasscharts}. Generally, they are consistent with the singularity at $\rho=0$ having a positive linear mass density equal to $-K/2$ at least for small negative values of $K$, while the outer singularity at $\rho=\rho_o$ has a negative linear mass density equal to $-K/2$ at least for small positive values of $K$. This is due to the fact that the Minkowski limit can only be applied in these regions and the corresponding signs of $K$, see the last paragraph of Chapter 3. If we shift the origin of coordinate $\rho$ to $\rho_o$ and calculate the energy within a cylinder centered at $\rho=\rho_o$, we find that the total mass switches sign.
\subsection{Charge}
We now calculate the charge within a closed area $S$
\begin{equation}
Q \equiv \frac{1}{4\pi} \oint_{S} *F = \frac{1}{4\pi} \oint_{S} F_{\alpha \beta} r^{\alpha} n^{\beta} \diff{S} = \frac{-hK}{2},
\end{equation}
where $F$ is the Maxwell tensor and $*$ denotes the Hodge dual. Similarly to the previous subsections on energy, if we calculate the charge enclosed by a cylindrical surface around the outer axis, the expression switches its sign. We thus conclude that the singularity at $\rho=0$ has a linear charge equal to $-K/2$ just like the outer singularity when observed from the outside. However, when we consider the outer singularity from below, its linear charge density is $K/2$. Overall, the linear charge and mass densities of the singularities are of the same magnitude as expected.
\section{Shell sources}
We now apply the Israel formalism \cite{israel} generalized to non-vacuum spacetimes involving electromagnetic fields by Kucha\v{r} \cite{kuchar}. Instead of relying on non-unique definitions of local energy density, we replace the singular regions of the spacetime by flat space and study the mass and charge of matter induced on the interface between the newly introduced Minkowski sections and the original ECS outside. For the sake of brevity, we just give the results here.

We first replace the singularity at $\rho=\rho_o$ and thus take $\rho>\rho_o$. We have a Minkowskian cylinder of finite radius restricted by an infinitely thin cylindrical surface of induced matter, beyond which ECS stretches on to radial infinity. For $K>0$ we must have $U>0$ and for the mass and charge induced per unit length of the cylindrical interface we find $M_1 = Q_1 = -\frac{1}{2} \frac{K}{U} < 0$. There is zero induced pressure along the $z$ and $\phi$ directions. The lowest-order expansion for the unit-length mass and charge yields $M_1=Q_1=-K/2$ as expected. Taking $K<0$ now, we have $U<0$ and the induced mass is of the form $M_1 = \frac{1}{2} (1+ \frac{K}{U}) > 0$ and the induced charge is $Q_1 = \frac{1}{2} \frac{K}{U} > 0$. We cannot apply the limit $K\rightarrow0$ here, since we could not keep the cylinder radius constant in the limiting process to stay above the outer singularity. Additionally, there is also induced tension along the $z$-direction.

Let us look now at the singularity located at $\rho=0$. For $K>0$ we need $U<0$ and find $M_1 = \frac{1}{2} (1+ \frac{K}{U})$ and $Q_1=\frac{1}{2}\frac{K}{U}<0$. There is again tension induced along the $z$ direction. The induced mass is positive below the radius of largest circumference, $\rho_c$ (see text below (\ref{hoop_circumference}) and \Fref{fig:properlength}), and negative above it. There is no Minkowski limit. Taking $K<0$ now, we need $U>0$ and the induced mass is of the form $M_1=Q_1=-\frac{1}{2}\frac{K}{U}>0$. The lowest-order expansion yields $M_1=Q_1=-K/2$ in the Minkowski limit as expected. The induced cylinder has no pressure or tension in any direction.

Finally, we now replace the outer singularity---located at $\rho_o$ and forming the other spacetime axis with a vanishing circumference (see \Fref{fig:properlength})---by a cylindrical region of Minkowski and continue across a cylindrical surface inside, towards $\rho=0$. Let us first take $K>0$ and thus $U<0$. We have $M_1=Q_1=-\frac{1}{2} \frac{K}{U} >0$, with no pressure or tension. We cannot do the Minkowski limit here. For $K<0$ and $U>0$ we obtain $Q_1 = \frac{1}{2} \frac{K}{U} <0$ and $M_1 = \frac{1}{2} (1+ \frac{K}{U})$, which is negative if we are above $\rho_c$. There is again tension along the $z$-direction. We cannot do the Minkowski limit since the cylinder radius would fall below $\rho_c$ and we would end up with two sections of Minkowski pasted together inside out with two axes present in the resulting spacetime.

The cases admitting a Minkowski limit are consistent with the findings of Chapter~4 and $|M_1|=|Q_1|=|K|/2$. For an overview, see Conclusions and summary.
\section{Equations of motion}
Finally, we will study motion of test particles in the spacetime to compare to the results from the previous sections. The Lagrangian for a charged particle of specific charge $q$ and moving in an electromagnetic field is
\begin{eqnarray}
\Lagr &=& \frac{1}{2}g_{\mu \nu}\dot{x}^{\mu}\dot{x}^{\nu}+q\dot{x}^{\kappa}A_{\kappa}=\\
\nonumber
&=&\frac{1}{2} \left[\left(\rho ^2 \dot{\phi }^2 +\dot{\rho }^2+\dot{z}^2 \right)U ^2 -\frac{\dot{t}^2}{U ^2}\right]+q\frac{\dot{t}}{U}.
\end{eqnarray}
The Lagrangian does not explicitly depend on $t, \phi$ and $z$, so the integrals of motion read
\begin{equation}
E \equiv \frac{q U -\dot{t}}{U ^2}, L_z \equiv \rho ^2 \dot{\phi } U ^2 , N \equiv \dot{z} U ^2.
\end{equation}
There is thus a single equation remaining, which is not explicitly integrated:
\begin{equation}
\ddot{\rho}-\rho  \dot{\phi}^2 - \frac{U_{,\rho}}{U } \left[ \rho ^2 \dot{\phi}^2-\dot{\rho}^2+\dot{z}^2\right]+\dot{t} U_{,\rho}\frac{q U -\dot{t}}{U^5}=0.
\end{equation}
And finally the normalization
\begin{equation}
\left(\rho ^2 \dot{\phi }^2 +\dot{\rho }^2 +\dot{z}^2 \right)U ^2 -\frac{\dot{t}^2}{U ^2} = \Unor,
\end{equation}
where $\Unor$ is a normalization constant, $\Unor = 0$ for photon motion and $\Unor = -1$ for timelike motion. The equations of motion are singular if $U$ or $U_{,\rho}$ are singular and they diverge at $\rho = 0$ or $\rho = \rho_o$.
\subsection{Static electrogeodesics}
First we will investigate static solutions. The equations reduce to
\begin{equation}
-\frac{\dot{t}^2}{U ^2} = -1, \ddot{t} = 0, \left(qU - \dot{t}\right)\dot{t} = 0.
\end{equation}
The solution only exists between the singularities for $q=-\sgn K$ and outside for $q=\sgn K$. Accepting our previous results on the charge of the singularities, we see that for $K>0$ a test particle located between the singularities with $q=-1$ is repelled outward so the linear mass density of the two singularities must be positive at $\rho=0$ and negative at $\rho=\rho_o$ at least in the weak-field limit. Moreover, their magnitude must be the same as this is the case for the particle. If we are outside where $q=1$, the particle is attracted inwards by the charge so it must be repelled by a negative mass density. Following the same line of reasoning for $K<0$, we find that the signs of mass densities of the singularities are in fact the same as for $K>0$ and their magnitude is again equal to the charge density.
\subsection{Radial motion}
Radial cylindrical electrogeodesics are defined as world-lines where $\phi$ and $z$ are independent of proper time. They are governed by the equations
\begin{eqnarray}
\label{eq:radial1}
\dot{\rho }^2 U ^2 -\frac{\dot{t}^2}{U ^2} &=& \Unor, \\
\label{eq:radial2}
\frac{q U -\dot{t}}{U ^2} &=& E,
\end{eqnarray}
where the first equation is the normalization condition and the second one comes from conservation of $E$.
\subsubsection{Photon motion}
Taking $q = \Unor = 0$, (\ref{eq:radial1})--(\ref{eq:radial2}) become equations of motion for photon. We proceed by expressing $\dot{t}$ from (\ref{eq:radial2}). From normalization we then obtain an equation for $\rho$. We thus need to solve
\begin{equation}\label{eq:radialphoton}
\dot{t} = -E \left(1+K \ln \rho\right)^2, \dot{\rho}^2 = E^2.
\end{equation}
From (\ref{eq:radialphoton}) we immediately see that $E < 0$ for $\dot{t}$ to be positive. For $E = 0$ the photon would be static. This solution reads
\begin{equation}
\rho (\tau) = r_0 \pm |E| \tau,
\end{equation}
where $r_0$ is the initial radius and $\tau$ is the affine parameter. If the photon starts towards one of the singularities, it will hit it, only photons emitted above outer singularity can avoid both singularities. The geodesic cannot be continued across the singularities and, therefore, we have two regions of spacetime which are causally separated.
\subsubsection{Electrogeodesic}
\begin{figure}[h]
\centering
\subfloat[$K<0$]{\includegraphics[width = 3in]{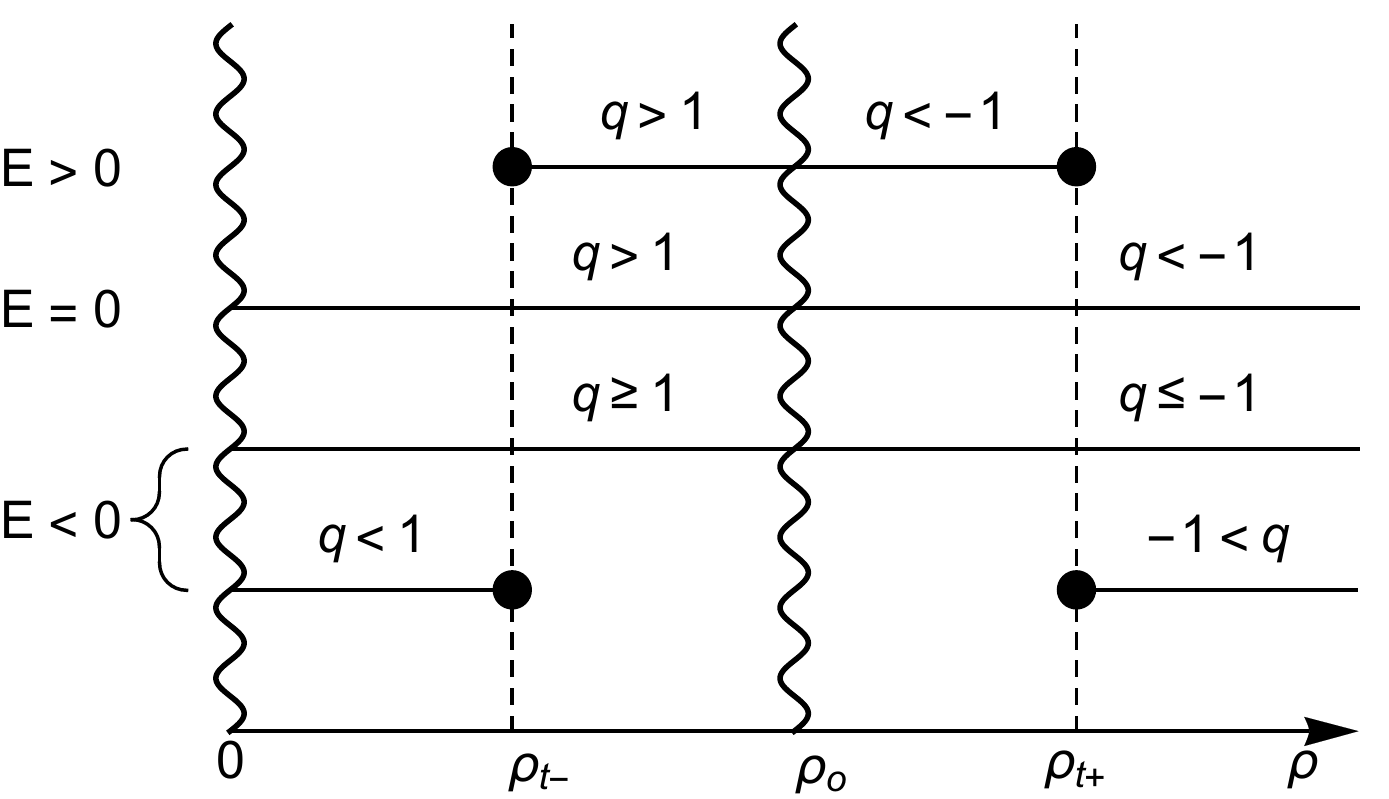}}
\subfloat[$K>0$]{\includegraphics[width = 3in]{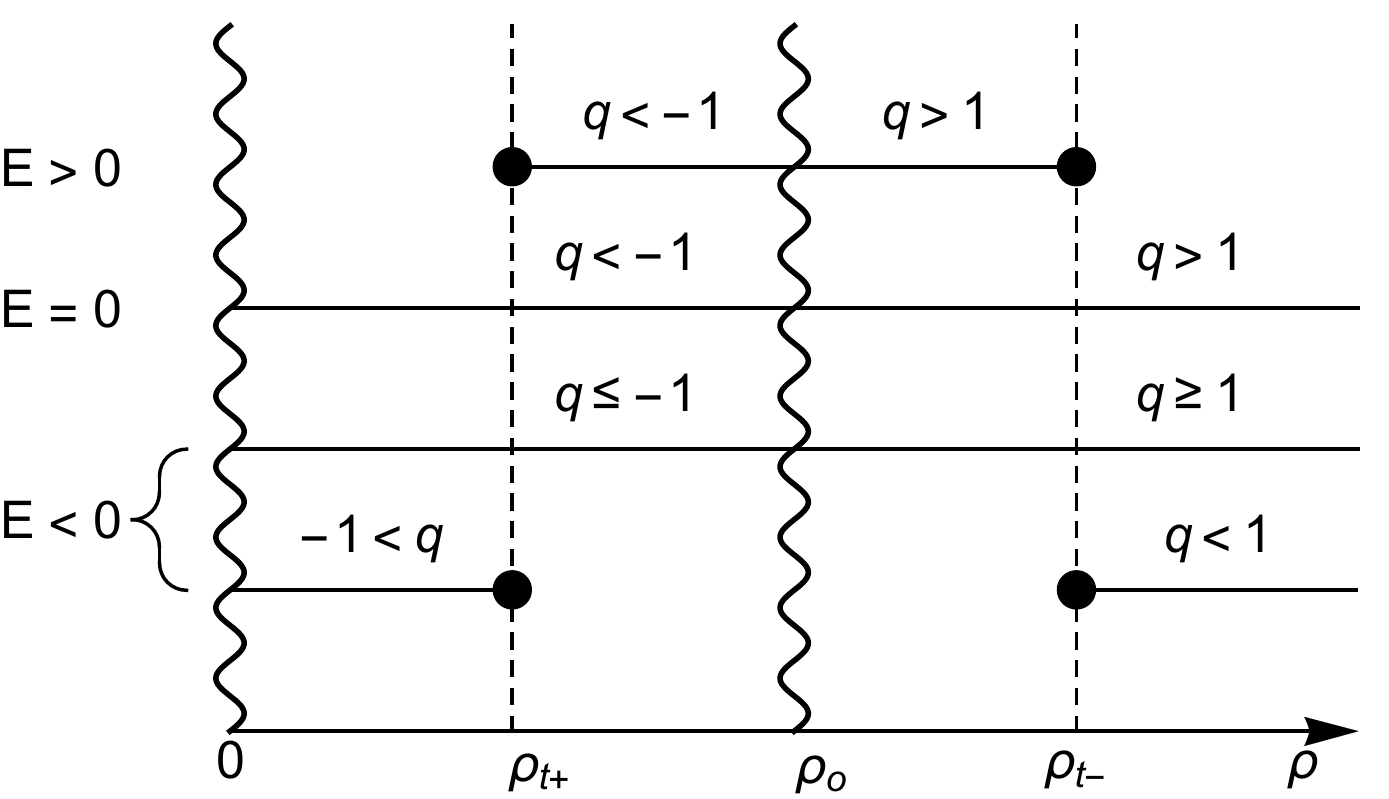}}\
\caption{Regions, where radial electrogeodetic motion is possible.}
\label{fig:Radial Electrogeodesics}
\end{figure}

The equations for electrogeodesic can be integrated to yield
\begin{eqnarray}
\dot{t} = - \left(1+K \ln \rho \right)\left(E-q+E K \ln \rho \right), \\
\dot{\rho}^2 = \frac{\left(E - q+ E K \ln \rho \right)^2 - 1}{\left(1+K \ln \rho \right)^2}.
\end{eqnarray}
We find two turning points
\begin{equation}
\rho_{t\pm} = \exp \frac{q-E\pm 1}{E K}, E \neq 0, q \neq \mp 1.
\end{equation}
Radial acceleration at these radii reads
\begin{equation}
\ddot{\rho} \left(\tau : \rho = \rho_{t\pm}\right) = \pm \frac{K E^3}{\rho_{t\pm} \left(q \pm 1\right)^2} \neq 0 \mathrm{~for~} q \neq \mp 1.
\end{equation}
We summarize our results for radial electrogeodesics in \Fref{fig:Radial Electrogeodesics}.
\subsection{Circular electrogeodesics}
We investigate circular electrogeodesics with $\rho$ and $z$ constant and governed by
\begin{eqnarray}
-\frac{\dot{t}^2}{U ^2}+U ^2\rho ^2 \dot{\phi} ^2  &=& \Unor,\\
\ddot{t} = \ddot{\phi} &=& 0 ,\\
\frac{q \dot{t} U_{,\rho}}{U^4}-\frac{\dot{t}^2 U_{,\rho}}{U^5}-\rho  \dot{\phi}^2-\frac{\rho ^2 \dot{\phi}^2 U_{,\rho}}{U} &=& 0.
\end{eqnarray}
We can immediately write $t = \gamma \tau , \phi = \omega \tau$ and insert this into the above equations to give
\begin{eqnarray}
-\frac{\gamma^2}{\left(1+K \ln \rho\right)^2}+ \left(1+K \ln \rho\right)^2 \rho ^2 \omega ^2  &=& \Unor, \\
 \frac{K \gamma ^2}{1+K \ln \rho}+ \rho ^2 \omega ^2 \left(1+K \ln \rho\right)^3\left(1+K+K \ln \rho \right) &=& q K \gamma.
\end{eqnarray}
\subsubsection{Photon motion}
Setting $q = \Unor = 0$ in the previous equations, we obtain
\begin{eqnarray}
\rho ^2 \omega ^2 U ^4 -\gamma ^2=0, \\
U_{,\rho} \gamma ^2+ U ^4 \rho \omega ^2 \left(U+\rho U_{,\rho}\right) &=& 0,
\end{eqnarray}
which yields the radius, $\rho_{ph}$, of the photon orbit
\begin{equation}
U+2K = 0 \Rightarrow \rho = \rho_{ph} \equiv e^{-2-1/K}, \gamma^2 = \rho ^2 \omega^2 K^2 ,
\end{equation}
where $\omega$ is a free parameter.
\subsubsection{Charged massive particle}
\begin{figure}[h]
\centering
\subfloat[Between singularities.]{\includegraphics[width = 3in]{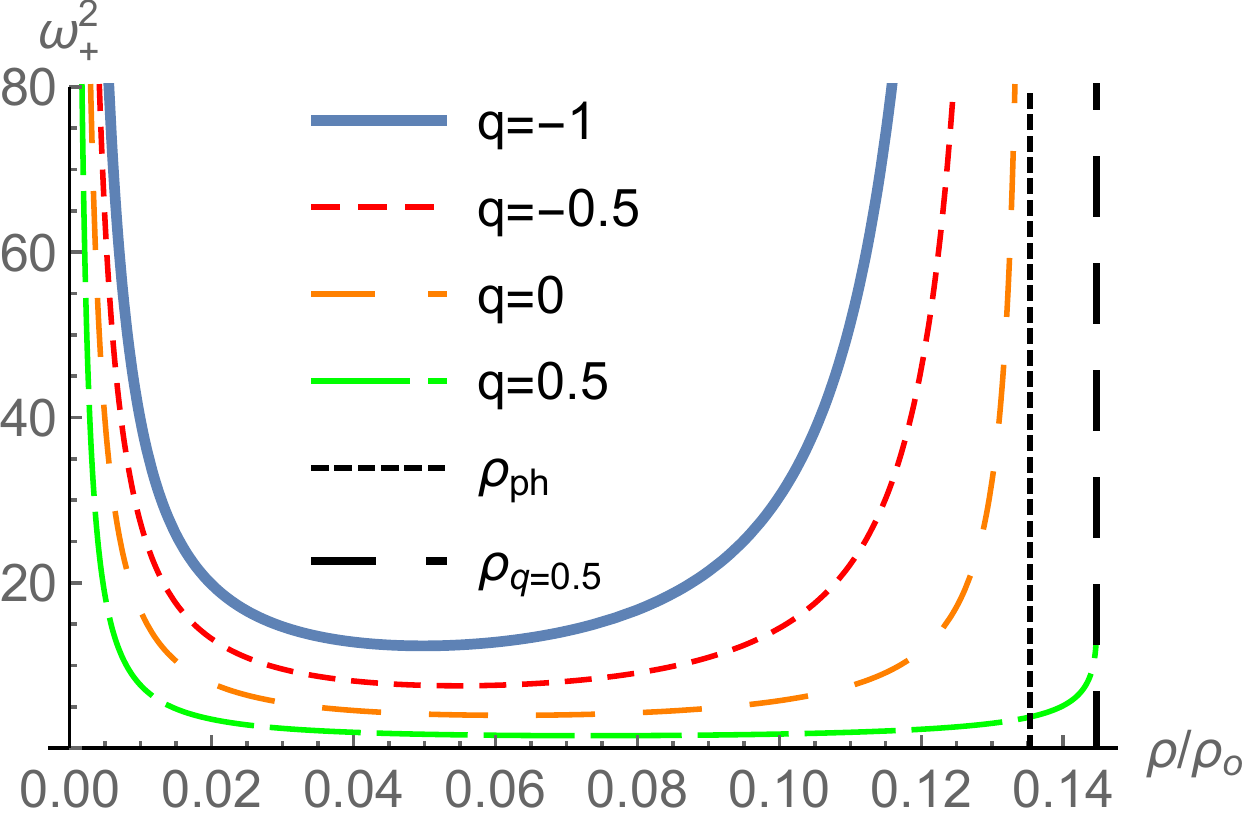}}
\subfloat[Outside.]{\includegraphics[width = 3in]{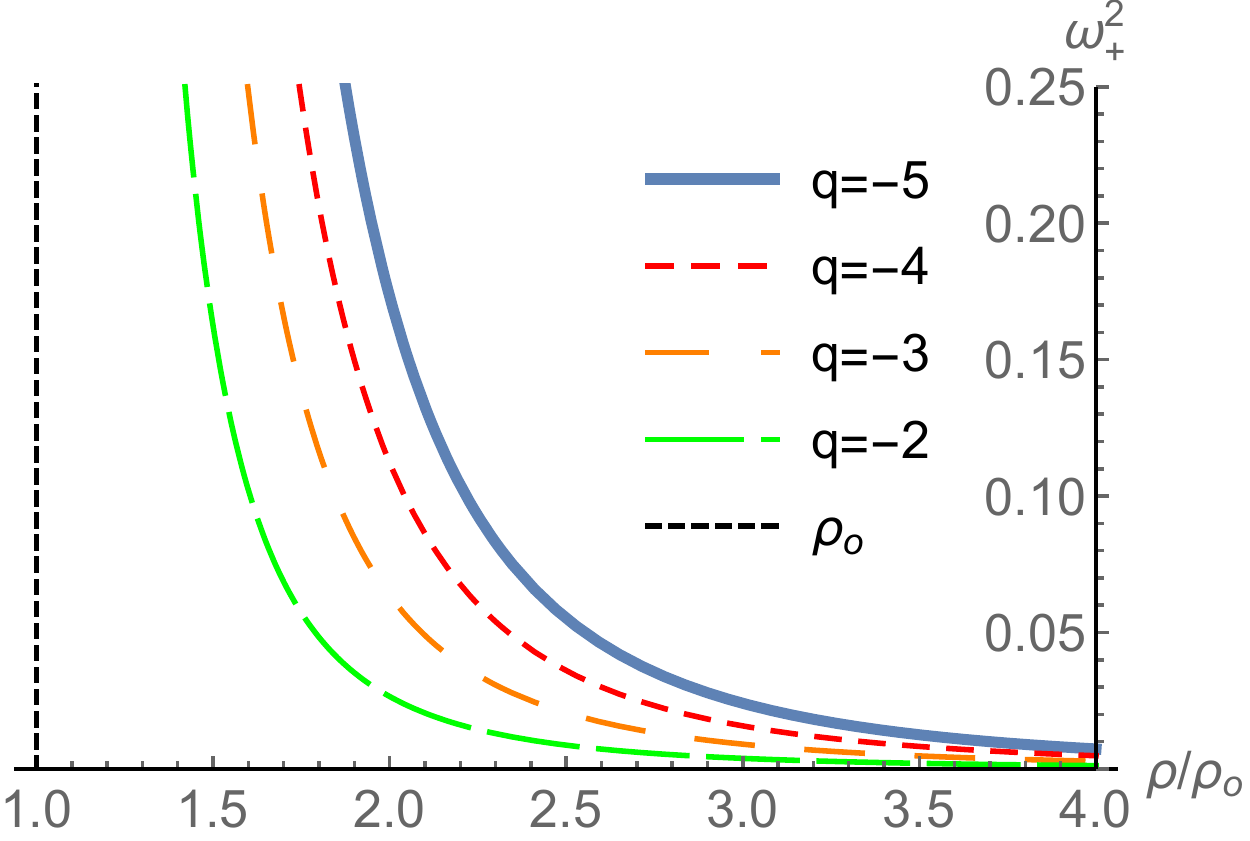}}\\
\subfloat[Between singularities.]{\includegraphics[width = 3in]{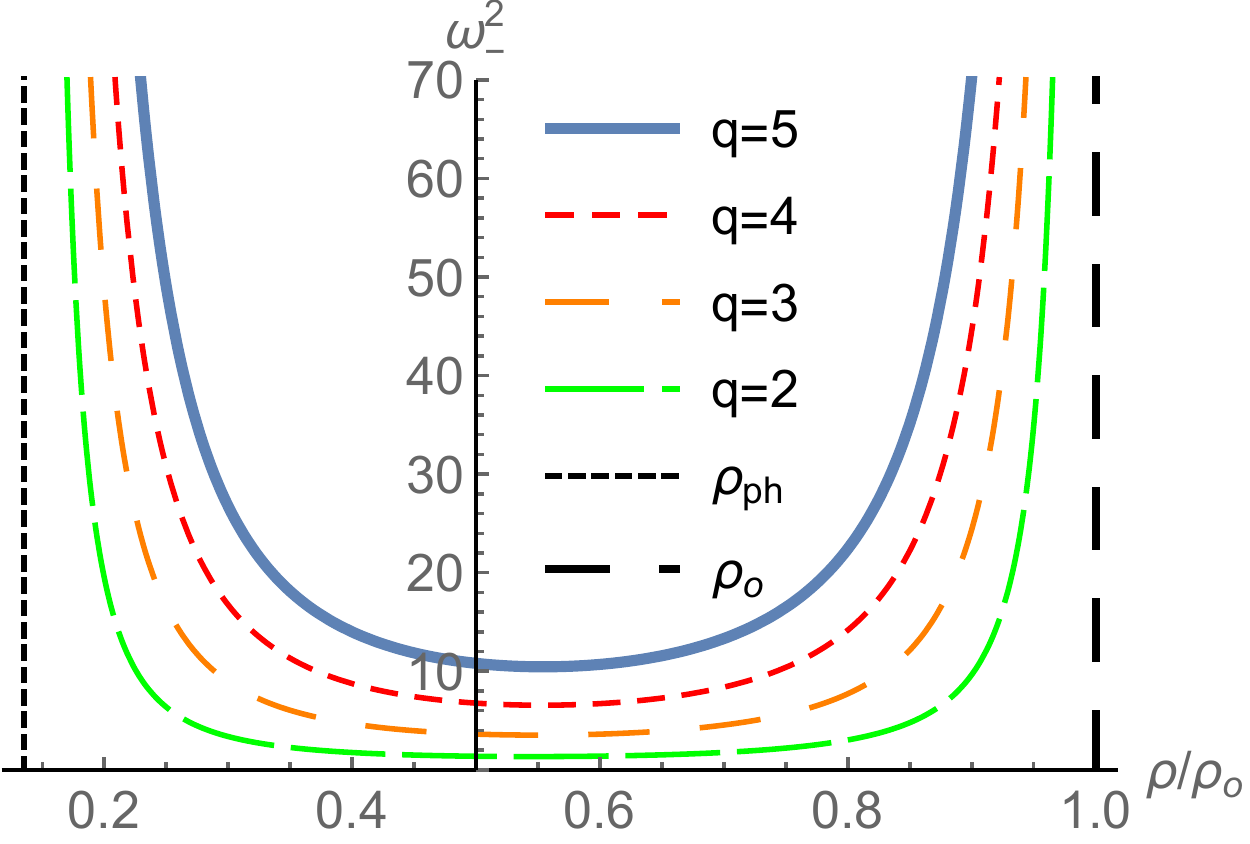}}
\subfloat[Between photon orbit and $\rho_q$.]{\includegraphics[width = 3in]{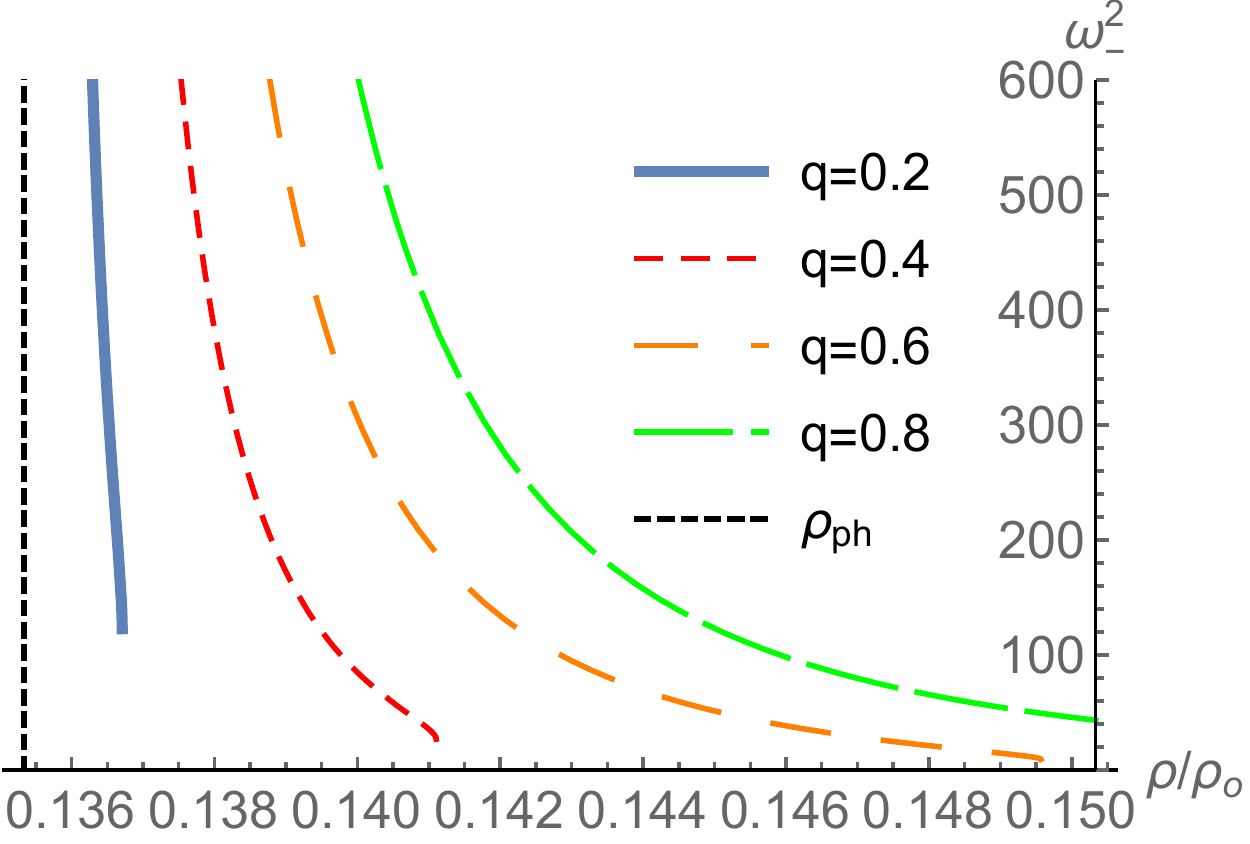}}\\
\caption{Plots of angular velocity for circular motion of test particles of varying specific charge as a function of the radial coordinate, $K=-2$.}
\label{circularelgeod}
\end{figure}

We now investigate charged massive particles with $\Unor = -1$. The equations are quadratic, so we expect two different absolute values of $\omega$ at most (the opposite sign corresponds to the opposite direction). First we express $\omega$ from the normalization condition
\begin{equation}\label{eq:gammanaomega}
\omega ^2 = \frac{\gamma ^2 - U^2}{\rho ^2 U^4}
\end{equation}
and substitute back into the second equation. The general solution (for $\rho \neq \rho_{ph}$) is
\begin{eqnarray}
\gamma_{\pm} &=& U\frac{ q \rho  U_{,\rho}\pm \sqrt{\left(q^2+8\right) \rho ^2 U_{,\rho}^2+12 \rho  U U_{,\rho}+4 U^2}}{2 \left(2 \rho  U_{,\rho}+U\right)},\\
\omega_{\pm}^2 &=&U_{,\rho}\frac{\rho U_{,\rho} \left(  q^2-4\right)-2 U \pm q \sqrt{\left(q^2+8\right) \rho ^2 U_{,\rho}^2+12 \rho  U U_{,\rho}+4 U^2}}{2 \rho  U^2 \left(2 \rho  U_{,\rho}+U\right)^2}.
\end{eqnarray}
Plots of the angular velocities for a range of specific charges are given in \Fref{circularelgeod}.

\begin{figure}[h]
\centering
\subfloat[$K<0$]{\includegraphics[width = 3in]{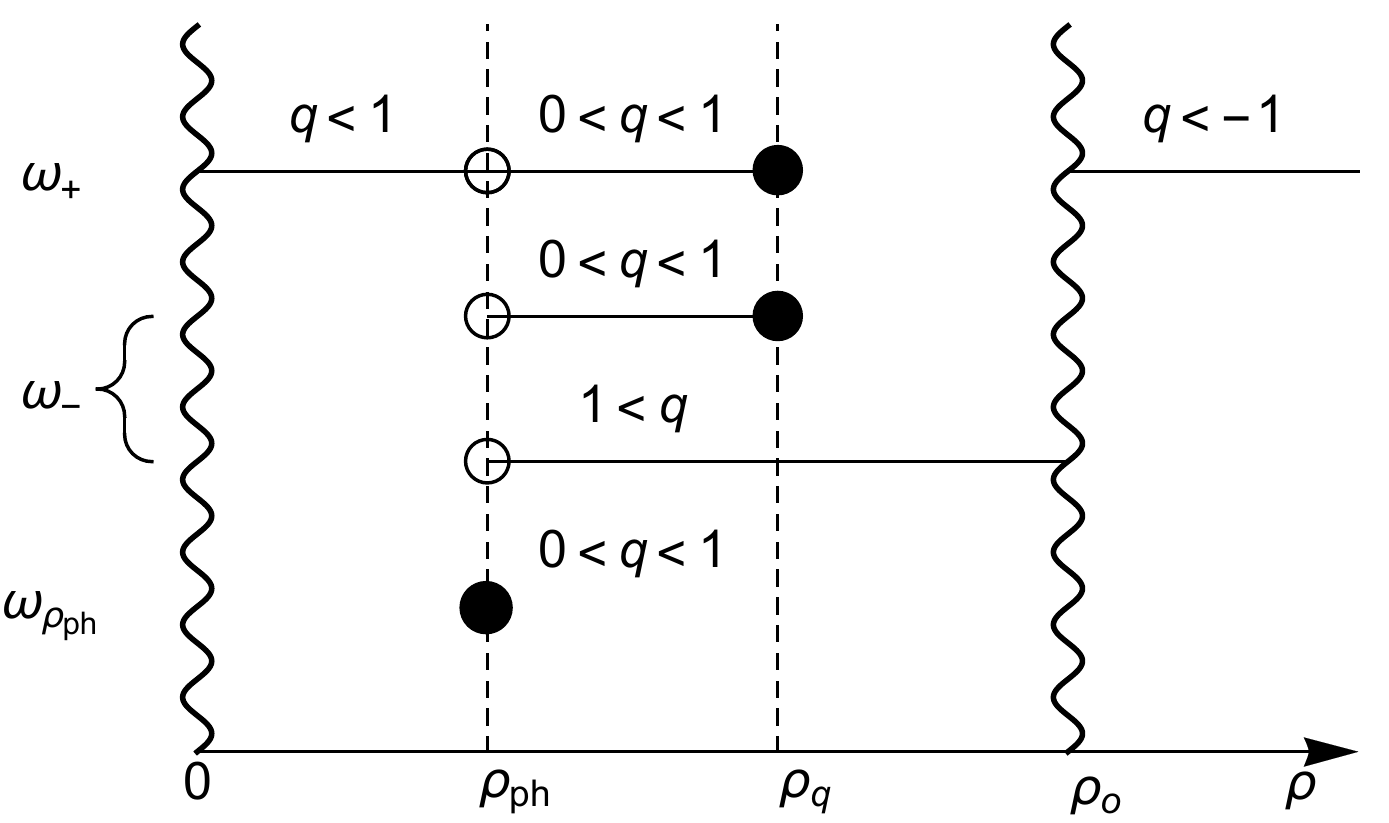}}
\subfloat[$K>0$]{\includegraphics[width = 3in]{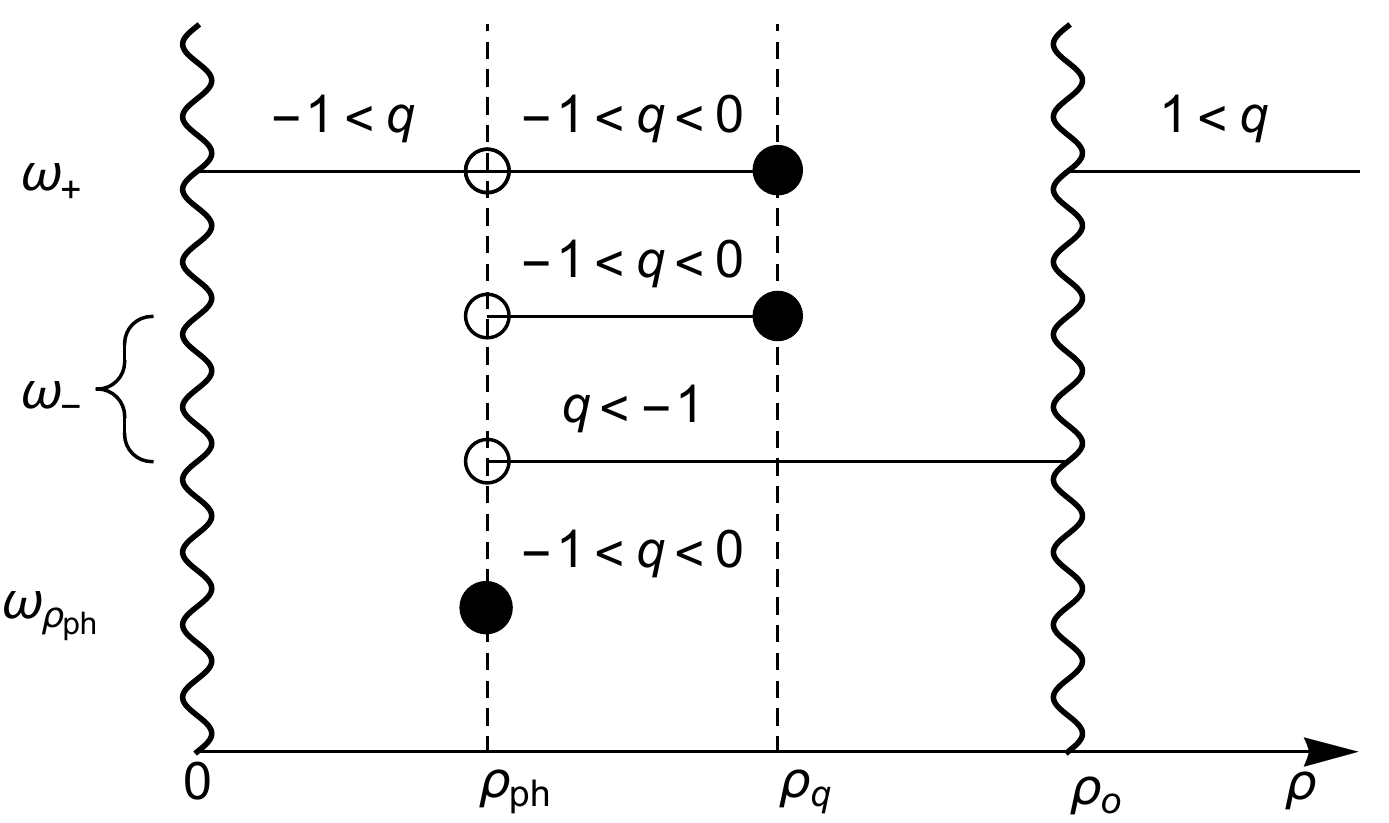}}\
\caption{Schematic illustration of regions where circular electrogeodesic motion is possible. The diagrams are not to scale, static solutions are excluded. The photon orbit radius $\rho_{ph}$ admits a special frequency $\omega_{\rho_{ph}}^2 = (1-q^2)/4 K^2 \rho_{ph}^2$.}
\label{circularelgeod_existence}
\end{figure}

To compare these trajectories to the Newtonian case, we calculate a series expansion of the angular velocity for $K \rightarrow 0$
\begin{equation}
\omega_{\pm}^2 \approx K \frac{-1\pm q}{\rho^2} + O\left(K^2\right).
\end{equation}
We need to set $\mu = \lambda$ in (\ref{eq:klasstatic}) to have an extremally charged source of the field and express the angular velocity, $\omega_N$, using the charge-to-mass ratio
\begin{equation}
\omega_{N}^2 = 2\lambda \frac{ 1  - q }{\rho ^2 }.
\end{equation}
Therefore, $\omega_+$ approaches the Newtonian velocity if we identify the parameter of the field with the unit-length mass and charge as follows
\begin{equation}
M_1 = Q_1 = -\frac{K}{2}.
\end{equation}
The regions where circular motion is possible are summarized in \Fref{circularelgeod_existence}. There are regions admitting both $\omega_\pm$ unlike the Newtonian case which only allows a single angular velocity at any given radius. This is a behavior we have already observed in a previous paper on another MP solution involving two charged black holes \cite{ryznerzofka} and is a result of the quadratic nature of the algebraic form of the equations of motion. The value $\rho_{q}$ used in the figures is
\begin{equation}
\rho_{q} = \exp \left(-\frac{2+3K+K \sqrt{1-q^2}}{2K}\right).
\end{equation}
\subsection{Electrogeodesics parallel to the axis}
\begin{figure}[h]
\centering
\includegraphics[width = 3in]{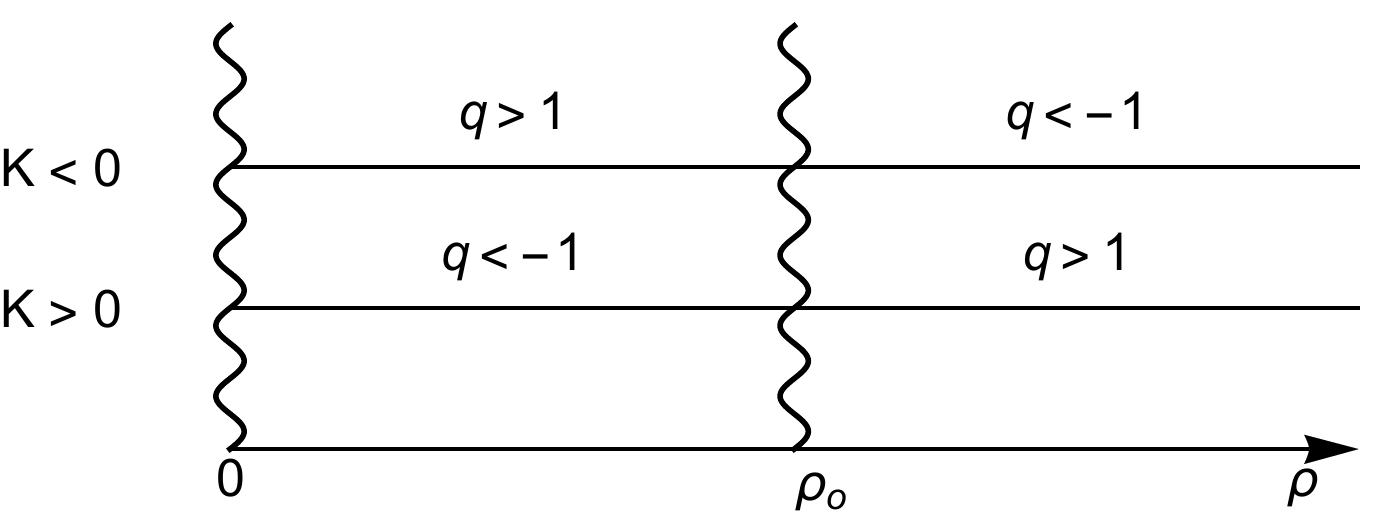}
\caption{Regions, where $z$-electrogeodetic motion is possible for $K \neq 0$}
\label{zelgeod}
\end{figure}

Unlike in the Newtonian case, motion parallel to the axis is possible in GR. It is governed by the following equations
\begin{eqnarray}
-\frac{\dot{t}^2}{U ^2}+\dot{z}^2 U^2 &=&\Unor,\\
\ddot{t} = \ddot{z} = 0 \Rightarrow t = \gamma \tau, z & = & v \tau, \\
U_{,\rho}\left(-q\dot{t}U +\dot{t}^2 + \dot{z}^2 U ^4 \rho \right) &=& 0.
\end{eqnarray}
Null geodesics parallel to the $z$-axis are not possible. For charged massive particles we determine their Lorentzian factor, $\gamma$, and their velocity parallel to the axis, $v$. We again find two distinct solutions
\begin{equation}
\gamma_\pm = \frac{U}{4} \left(q \pm \sqrt{q^2+8} \right), v_\pm = \sqrt{\frac{q^2 \pm q\sqrt{q^2+8}-4}{8 U ^2 }}.
\end{equation}
The regions admitting solutions are summarized in \Fref{zelgeod}.
\section{Conclusions and summary}
We investigated the physical properties of a cylindrically symmetric Majumdar-Papapetrou solution of Einstein-Maxwell equations sourced by a non-compact, extremally charged linear singularity forming the axis of the spacetime. Based on the form of the metric, we discovered that in addition to the axis singularity, the spacetime includes another singularity of different physical properties, which divides the spacetime into two causally separated regions. Inside, the situation is complicated by the presence of two singularities pulling and pushing test particles in opposite directions, while outside there is only one singularity present in the spacetime. Since ECS is a solution of electro-vacuum equations both singularities are the only source of the resulting gravitational and electromagnetic fields. Our goal was to establish their physical parameters---i.e., mass and charge densities per unit length---and geometric characteristics. In this respect, we found that $|M_1|=|Q_1|=|K|/2$, see below.

We calculated motion of charged test particles moving in preferred directions and compared it to the solution within the framework of classical mechanics. Based on the behavior of radial electrogeodesics, we found that the singularities are not covered by horizons and are thus naked in accordance with \cite{hartle}. Next, we dealt with static trajectories and gained insight into the specific charge of the singularities. We then determined circular paths and regions where such motion is possible. One interesting result is a range of circular orbit radii and conditions on the spacetime parameters allowing the existence of two electrogeodesics of the same radius but differing angular velocities in the same direction. We then determined the classical limit of the circular velocities to again obtain information about the specific charge of the singularities.

When determining the mass of the sources of ECS, we also proceeded from the total energy of a region of spacetime containing the singularity. However, there is generally no way of determining locally the energy of the gravitational field in GR and we thus used several different definitions and compared them in the weak-field regime where they all agreed.

Using Israel formalism, we then regularized the spacetime by replacing the regions containing the singularities with flat space and continuing with ECS outside the respective cylindrical hypersurfaces, which then become the source of the gravitational and electromagnetic fields. We determined their charge and mass and compared them again for various spacetime parameters and also with the classical solution.

\begin{figure}[h]
\centering
\subfloat[$K<0$]{\includegraphics[width = 3in]{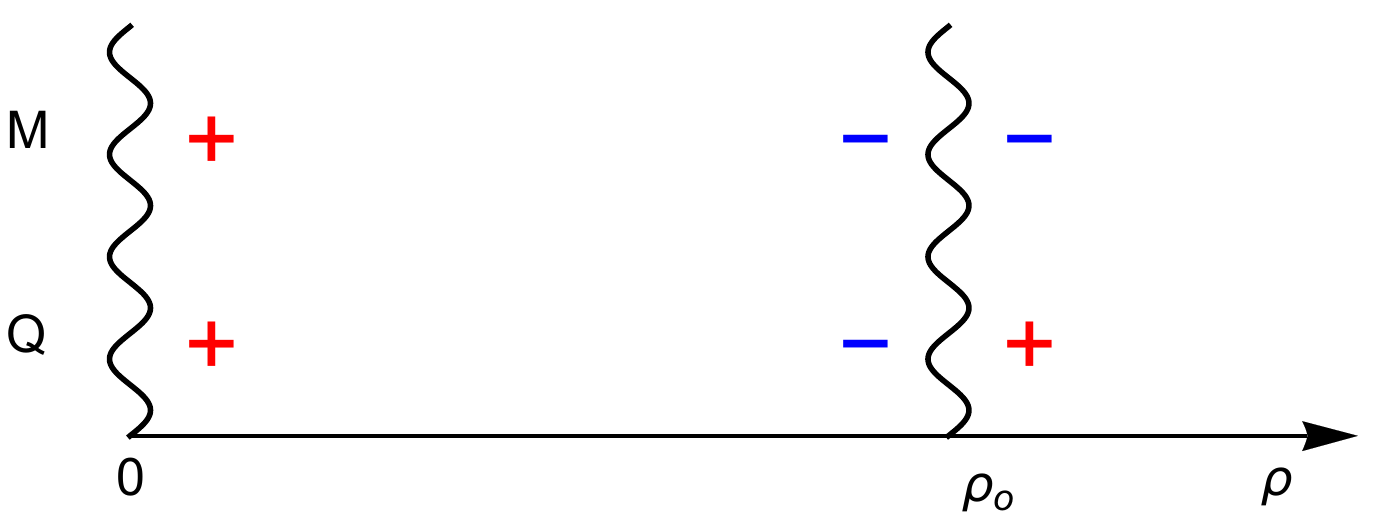}}
\subfloat[$K>0$]{\includegraphics[width = 3in]{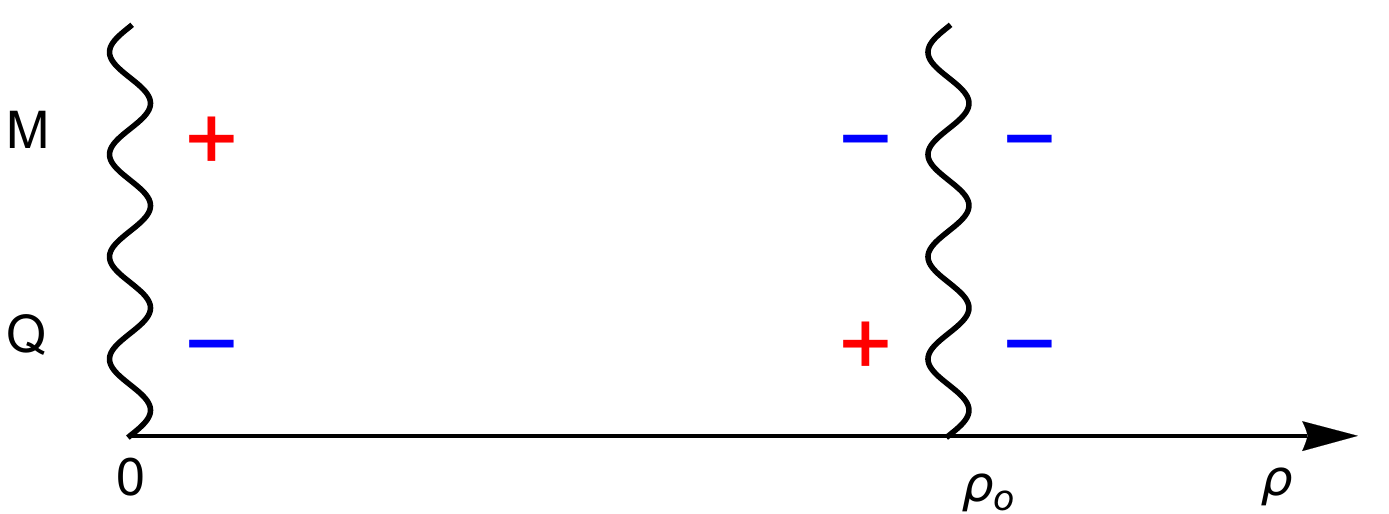}}\
\caption{Schematic illustration of the sign of linear mass and charge densities of the singularities. Their magnitude is always equal to $|K|/2$, see our conclusions.}
\label{sign_of_linear_densities}
\end{figure}

We finally conclude that the parameter $K$ appearing in the metric determines both mass and charge of the singularities per unit length, $M_1$ and $Q_1$, respectively. For weak fields, we conclude that $|M_1|=|Q_1|=|K|/2$ as expected since the MP sources should be extremal. The inner singularity at $\rho=0$ has an electric charge per unit length $Q_1=-K/2$ and it always has a positive mass per unit length $M_1=|K|/2$. The outer singularity at $\rho=\rho_0$ always has a negative mass per unit length, $M_1=-|K|/2$, as observed from both inside and outside, while its charge per unit length is $Q_1=K/2$ observed from the inside and $Q_1=-K/2$ observed from the outside. See \Fref{sign_of_linear_densities} for the signs. It is, however, difficult to give a completely general interpretation of $K$ as we need to build our intuition based on comparisons with situations we understand. That is similar, for instance, to the Levi-Civita case where the local parameter of the metric---denoted $m$ or, sometimes, $\sigma$---determines mass per unit length of the source only for small values.

Using several independent methods, we thus clarified the meaning of the spacetime's structure and parameter and gained intuition about its physical interpretation. The results will broaden the knowledge of MP spacetimes with non-compact sources, which have not been paid much attention so far.
\section*{Acknowledgements}
JR was supported by Student Faculty Grant of Faculty of Mathematics and Physics, Charles University in Prague and by grant No. 504616 of Charles University Grant Agency. MZ was supported by Albert Einstein Center, Project of Excellence No. 14-37086G funded by the Czech Science Foundation.

\end{document}